\documentclass[prb,twocolumn,amsmath,amssymb]{revtex4}
\usepackage{graphicx}
\usepackage{dcolumn}
\usepackage{bm}

\begin{document}



\title{Kramers-Kronig constrained variational analysis of optical spectra}

\author{A. B. Kuzmenko}

\affiliation{DPMC, University of Geneva, 24 Quai Ernest-Ansermet,
1211 Geneva 4, Switzerland\\}

\begin{abstract}
A universal method of extraction of the complex dielectric
function
$\epsilon(\omega)=\epsilon_{1}(\omega)+i\epsilon_{2}(\omega)$ from
experimentally accessible optical quantities is developed. The
central idea is that $\epsilon_{2}(\omega)$ is parameterized
independently at each node of a properly chosen anchor frequency
mesh, while $\epsilon_{1}(\omega)$ is dynamically coupled to
$\epsilon_{2}(\omega)$ by the Kramers-Kronig (KK) transformation.
This approach can be regarded as a limiting case of the
multi-oscillator fitting of spectra, when the number of
oscillators is of the order of the number of experimental points.
In the case of the normal-incidence reflectivity from a
semi-infinite isotropic sample the new method gives essentially
the same result as the conventional KK transformation of
reflectivity. In contrast to the conventional approaches, the
proposed technique is applicable, without readaptation, to
virtually all types of linear-response optical measurements, or
arbitrary combinations of measurements, such as reflectivity,
transmission, ellipsometry {\it etc.}, done on different types of
samples, including thin films and anisotropic crystals.

\end{abstract}

\maketitle

\section{Introduction}

The linear macroscopic optical characteristics of materials, such
as reflectivity, absorption, penetration depth and others, are
largely determined by the complex dielectric function
$\epsilon(\omega)=\epsilon_{1}(\omega) + i\epsilon_{2}(\omega)$
\cite{Text1}. Typically, one has to extract $\epsilon_{1}(\omega)$
and $\epsilon_{2}(\omega)$ of the studied sample by the inversion
of measured optical spectra. These quantities are not independent,
but comply the Kramers-Kronig (KK) relations
\cite{KramersNature26,KronigJOSA26,LandauLifshitz,DresselGruner}:
\begin{equation}\label{kkeps1}
\epsilon_{1}(\omega) - 1 =
\frac{2}{\pi}\wp\int_{0}^{\infty}\frac{x\epsilon_{2}(x)}{x^2-\omega^2}dx,
\end{equation}
\begin{equation}\label{kkeps2}
\epsilon_{2}(\omega) =
-\frac{2\omega}{\pi}\wp\int_{0}^{\infty}\frac{\epsilon_{1}(x)}{x^2-\omega^2}dx
+\frac{4\pi\sigma_{DC}}{\omega},
\end{equation}
\noindent where $\sigma_{DC}$ is the DC conductivity, and the
symbol $\wp$ denotes the principal value integral.

When talking about experimental ways to extract of $\epsilon_{1}$
and $\epsilon_{2}$, one should distinguish between the ones that
rely on the KK relations and the ones that allow the determination
of both quantities directly from the experiment. A well-known
example of the first type of techniques is the KK analysis of the
normal-incidence reflectivity $R(\omega)$ spectra
\cite{JahodaPRB57}. The idea of this method is the following.
Equations (\ref{kkeps1}) and (\ref{kkeps2}) cannot be directly
applied, since both $\epsilon_{1}$ and $\epsilon_{2}$ depend on
the unknown phase $\theta$ of the complex reflectivity
\begin{equation}\label{r}
r =
\frac{1-\sqrt{\epsilon}}{1+\sqrt{\epsilon}}=\sqrt{R}\exp{(i\theta)}.
\end{equation}

\noindent A very helpful 'trick' is to write down a relation
similar to (\ref{kkeps2}) with respect to the complex function
$\ln r(\omega)=\ln \sqrt{R(\omega)} + i\theta(\omega)$:
\begin{eqnarray}\label{kkr}
\theta(\omega) = -\frac{2\omega}{\pi}\wp\int_{0}^{\infty}\frac{\ln
\sqrt{R(x)}}{x^2-\omega^2}dx + \theta(0)
\end{eqnarray}
\noindent and use it to compute $\theta(\omega)$ from $R(\omega)$.
The reflectivity has to be extrapolated outside the experimentally
accessible frequency range, which may give rise to a large
uncertainty, if the experimental range is not sufficiently broad
\cite{AspnesBook}. Having found the reflectivity phase, one can
restore the dielectric function by inverting (\ref{r}): $\epsilon
= (1-r)^2/(1+r)^2$. Thus, the KK relation is an additional
condition, which 'miraculously' helps to get two spectra,
$\epsilon_{1}(\omega)$ and $\epsilon_{2}(\omega)$, out of one,
$R(\omega)$.

Unfortunately, the approach based on equations like (\ref{kkr}),
is not universal \cite{ChambersIP75}. The KK relations are derived
for the so-called response functions \cite{LandauLifshitz}, which
have regular analytical behavior in the upper complex frequency
half-plane due to the causality principle. In the case of the
function $\ln r(\omega)$, a necessary condition is that $r(\omega)
\neq 0$ for $\mbox{Im}(\omega)>0$. One can show, that it holds for
the normal angle of incidence, but breaks down, for instance, for
off-normal reflectivity, if the angle of incidence is large enough
\cite{SomalThesis}. The same problem appears, when the sample is a
thin film on a substrate and in many other cases. A number of
analytical and numerical algorithms were put forward to extend the
KK method to off-normal reflection
\cite{HaleAO73,MakarovaOS76,AlperovichOS79,BardwellJCP85,GrosseAPSS91,HopfeJPAP92,SomalThesis},
transmission
\cite{NeufeldJOSA72,BringansJPAP77,YamamotoVS97,VarsamisAS02},
attenuated total reflection (ATR) spectra
\cite{TshmelSA73,BardwellJCP85,YamamotoVS94,BertieJCP96} as well
as to optical data on layered samples
\cite{LupashkoOS70,PliethSS75,BringansJPAP77,GrosseAPSS91,TikhonravovMUPB93,SantanderPRL02,SantanderThesis}
and low-symmetry crystals \cite{KochCP74,KuzmenkoOS97} and other
'non-standard' situations. Although the proposed techniques, taken
together, certainly cover a substantial scope of applications, it
is desirable, from the practical point of view, to have a single
universal approach, which can be used in most of the currently
used and future optical experiments without major readaptation.

The least-square fitting of spectra using simple model dielectric
functions, such as Drude-Lorentz oscillators, also called the
dispersion analysis \cite{KleinmanPR61}, is practically more
robust \cite{EbPRL01,SantanderPRL02} than the mentioned above
KK-based techniques. However, the range of considered dielectric
functions is artificially restricted to a certain model, which
makes it hard to recover, without distortions, all significant
spectral details from the experimental data. On the other hand,
the introduction of new parameters to a model (for example, the
addition of extra Lorentz oscillators), makes the functional form
for the dielectric function more 'flexible' and, in a sense, less
model-dependent.

The method of the {\it KK-constrained variational fitting},
developed in this paper, is a 'hybrid' of the two approaches. It
inherits the robustness of the least-square fitting and the
ability of the KK analysis to extract the full spectral
information. In simple terms, it is a limiting case of the
multi-oscillator modelling, when the number of oscillators is so
large (of the order of the number of experimental points), that
the analysis becomes essentially {\it model-independent}. Since
the KK relations (\ref{kkeps1},\ref{kkeps2}) are automatically
preserved in the used functional form of the dielectric function,
the proposed approach can also be regarded, as an extension of the
standard KK methods.

The basic idea of the KK-constrained variational fitting was
already introduced in Ref.\cite{KuzmenkoOS97} for the special case
of the normal-incidence reflectivity of low-symmetry (monoclinic
and triclinic) crystals, where the standard KK method fails. The
KK relations for the diagonal and off-diagonal components of the
reflectivity tensor were used, which limited the applicability of
the technique. Here we overcome this limitation by referring
directly to the KK relations for the dielectric function.

The rest of the manuscript is organized as follows. In Section
\ref{Problem} we outline the general problem targeted by the
method. Section \ref{Fitting} discusses the variational fitting as
a limiting case of the usual data fitting with a large number of
oscillators. In Section \ref{Method} we describe the technique in
details. Section \ref{Examples} contains a number of application
examples. Finally, in Section \ref{Summary} we summarize the
results and overview the possible future developments.

\section{Problem formulation}\label{Problem}

We begin with a formal problem definition. Suppose, a set of
experimental datapoints $\{\omega_{i}^{exp}, q_{i}^{exp}, \delta
q_{i}^{exp}\}$ ($i$=1,...,$N_{exp}$) is given, where
$\omega_{i}^{exp}$ is frequency, $q_{i}^{exp}$ is an optical
quantity of a certain type (reflectivity, transmission,
penetration depth, an ellipsometric angle {\it etc.}) and $\delta
q_{i}^{exp}$ is an experimental error bar of $q_{i}^{exp}$. We
assume, that all optical quantities can be expressed via real and
imaginary parts of the material dielectric function
$\epsilon(\omega)$:
\begin{equation}\label{qcalc}
q_{i}^{calc}=f_{i}\left(\epsilon_{1}(\omega_{i}^{exp}),
\epsilon_{2}(\omega_{i}^{exp})\right).
\end{equation}
\noindent Different datapoints can be of different types,
therefore we keep the index $i$ in $f_{i}$, Using the
transformation (\ref{kkeps1}) (which can be symbolically written
as $\{\epsilon_{1}\}-1=KK(\{\epsilon_{2}\})$), we get:
\begin{equation}\label{qcalc1}
q_{i}^{calc}(\{\epsilon_{2}\}) =
f_{i}(1+KK\{\epsilon_{2}\}(\omega_{i}^{exp}),
\epsilon_{2}(\omega_{i}^{exp})).
\end{equation}
\noindent Note that $q^{calc}_{i}$ depends not only on
$\epsilon_{2}$ at $\omega_{i}^{exp}$, but also on
$\epsilon_{2}(\omega)$ at all other frequencies.

It is natural to look for {\it a dielectric function
$\epsilon(\omega)$, which satisfies the KK relations and provides
the best fit of all experimental datapoints}:
\begin{equation}\label{chisq}
\chi^2(\{\epsilon_{2}\}) \equiv \sum_{i=1}^{N_{exp}}
\left(\frac{q_{i}^{calc}(\{\epsilon_{2}\}) - q_{i}^{exp}}{\delta
q_{i}^{exp}}\right)^2\rightarrow \min.
\end{equation}
\noindent This relation formalizes the least-square approach,
which is justified in the case of the normal error distribution
\cite{NumericalRecipes}.

The minimization problem (\ref{chisq}) is not fully defined unless
we specify the functional space for $\epsilon_{2}(\omega)$. In
principle, one may consider all physically acceptable functions
$\epsilon_{2}(\omega)$ ($0 \leq \omega < \infty$), which are
non-negative and integrable ({\it i.e.} compatible with the f-sum
rule \cite{LandauLifshitz,DresselGruner}):
\begin{equation}
\int_{0}^{\infty}x\epsilon_{2}(x)dx < \infty.
\end{equation}
\noindent However, in such a broad functional space the
minimization problem becomes {\it ill-posed}
\cite{DienstfreyIP01}. In particular, multiple solutions for
$\epsilon_{2}(\omega)$ may exist, which look very different, but
minimize (\ref{chisq}) equally well. However, if the set of
datapoints is sufficiently broad and dense, only one (or maybe
few) solutions will show a regular behavior, while the majority of
them will be marked by 'suspicious' features, for instance, sharp
spikes, discontinuities or oscillations with a period smaller than
the distance between experimental frequencies.

It suggests that a proper way to regularize this ill-posed problem
is to restrict somehow the functional space
$\epsilon_{2}(\omega)$. It can be done using some parametrization,
for example, the decomposition of $\epsilon_{2}(\omega)$ in a
finite basis of fixed reference functions. An optimal
parametrization should exclude the mentioned irregular features,
while being flexible enough to fit all important spectral details.
Without rigorous mathematical proof, but rather based on practical
tests, we argue that {\it the optimal number of parameters should
be of the order of, but not larger than, $N_{exp}$}.

One should note that it is more common to use model dielectric
functions with a small number of parameters, much less than
$N_{exp}$. In this case, the purpose of the analysis is to get a
{\it physical insight} using the measured spectra directly. For
instance, the validity of a certain hypothesis about the studied
compound can be tested and the values of some meaningful
parameters (such as the plasma frequency, optical transition
energies {\it etc.}) can be found. On the contrary, the purpose of
parametrization in the {\it model-independent} method, proposed in
this paper, is not to get the values of parameters (which may have
almost no physical meaning, taken alone), but rather to extract
the complex dielectric function by the inversion of the available
data. Thus, it is not surprising that the number of parameters
becomes very large. The minimization of (\ref{chisq}) in this case
is essentially a {\it variational} problem. The obtained in this
way dielectric function can be used later on for a more specific
examination.

\section{Drude-Lorentz fitting as a parametrization}\label{Fitting}

An important example of parametrization of dielectric functions is
the Drude-Lorentz (DL) oscillators formula
\begin{equation}\label{DL}
\epsilon(\omega) = \epsilon_{\infty} +
\sum_{k=1}^{N}\frac{\omega_{p,k}^2}{\omega_{0,k}^2 - \omega^2 -
i\omega \gamma_{k}}.
\end{equation}
\noindent Each oscillator is represented by a Lorentzian with
three adjustable parameters: the oscillator frequency
$\omega_{0k}$, the linewidth $\gamma_{k}$ and the plasma frequency
$\omega_{p,k}$. The unbound oscillators (conductivity electrons)
are described by Drude terms, where $\omega_{0,k} = 0$. The
parameter $\epsilon_{\infty}$ is the contribution from the
higher-frequency oscillators.

Usually the number of oscillators is very limited, compared to
$N_{exp}$. As a result, the obtained spectra of $\epsilon(\omega)$
are 'unrealistically' smooth with some subtle, but potentially
important, spectral information being lost. According to previous
arguments, on should take a very large number of oscillators (of
the order of $N_{exp}$) in order to extract the information
model-independently. It was argued (see
Refs.\cite{EbPRL01,SantanderPRL02}) that the fitting of the most
important details of spectra (other than statistical noise) is not
much different from the KK transformation. The physical meaning of
individual oscillator terms may not be always clear, however, the
meaningful result is the complex value of $\epsilon(\omega)$
itself.

In practice, the oscillators are usually introduced one by one and
all Lorentzian parameters are allowed to change. Although this
strategy works very well for small N ($\sim$ 10 - 20), it becomes
cumbersome and badly controlled as the number of oscillators grows
further. In particular, as one adds more and more Lorentzians to
the model function (\ref{DL}), it becomes less and less clear how
to guess the initial parameter values. If all parameters are free
to change, it often happens that the fitting process becomes
numerically unstable. To avoid it, one has to deliberately fix
some parameters. It makes the fitting procedure rather tricky and
ambiguous, preventing the routinely operation. In the next section
we elaborate another fitting scheme which is more attractive from
the practical point of view.

\section{Method}\label{Method}

Now we proceed with the detailed description of the KK-constrained
variational fitting algorithm. Although it can be concisely
described in a hand-book like manner, we prefer to show its
'evolution' starting from the described above DL fitting scheme.
In this way, one can easily see that the new technique is, in
principle, a limiting case of the multi-oscillator fitting.

\subsection{Setting up the oscillators}

Let us consider a dense enough mesh of anchor frequency points
$\omega_{1}$,..., $\omega_{N}$. It must not the same as the set of
experimental frequencies $\omega_{1}^{exp}$,...,
$\omega_{N_{exp}}^{exp}$, introduced in Section \ref{Problem},
although they may coincide. We shall discuss below the optimal
choice of the anchor mesh.

\begin{figure}[thb]
\includegraphics[width=7cm,clip=true]{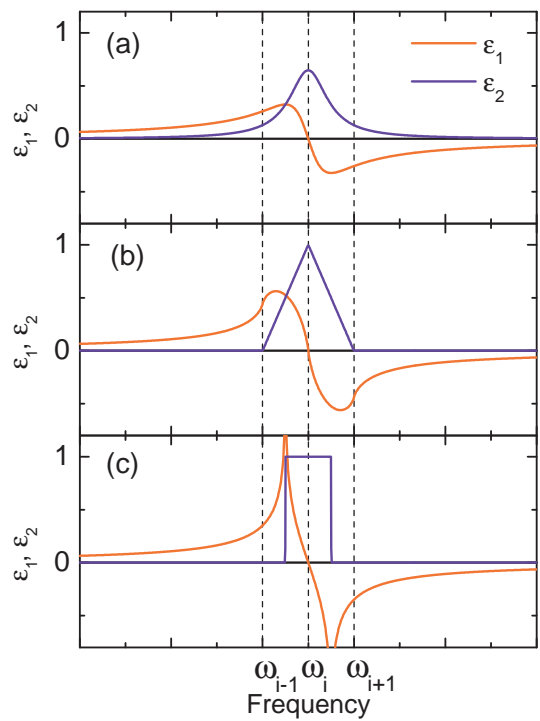}
\caption{Dielectric functions with a Lorentz (a), a triangular (b)
and a rectangular (c) lineshapes, centered at $\omega_{i}$ and
having the linewidth $\gamma_{i}=(\omega_{i+1}-\omega_{i-1})/2$.}
\label{FigShape}
\end{figure}

It is possible to set the number of oscillators in (\ref{DL}),
their frequencies  and linewidths in a rather logical and simple
way. To be 'flexible enough', the fitting function (\ref{DL})
should be able to deliver spectral weight {\it everywhere} in the
spectral range It can be done by putting one separate oscillator
{\it at every anchor frequency}: $\omega_{0,i} = \omega_{i}$.
Furthermore, we would like the $i$-th oscillator to be narrow and
provide spectral weight only in the closest neighborhood of
$\omega_{i}$. It suggests to set its width to a fixed value of the
order of the distance between adjacent frequency points, for
instance $\gamma_{i}=(\omega_{i+1}-\omega_{i-1})/2$. Figure
\ref{FigShape}(a) shows the real and the imaginary part of the
$i$-th Lorentzian:
\begin{eqnarray}\label{Lor}
\epsilon_{1,i}^{Lor}(\omega)=\frac{\omega_{p,i}^{2}(\omega_{i}^{2}-\omega^{2})}{(\omega_{i}^{2}-\omega^{2})^{2}+\gamma_{i}^{2}\omega^{2}}\\
\epsilon_{2,i}^{Lor}(\omega)=\frac{\omega_{p,i}^{2}\omega\gamma_{i}}{(\omega_{i}^{2}-\omega^{2})^{2}+\gamma_{i}^{2}\omega^{2}}
\end{eqnarray}

In this way we get a dense 'forest' of narrow Lorentzians, sitting
at fixed anchor frequencies. Only the oscillator strengths $S_{i}$
are left adjustable in order to set up the profile of the
'forest'.

\subsection{Optimizing the oscillator lineshape}

Because of extended low- and high-frequency tails, which interfere
with other oscillators, the Lorentz lineshape (Figure
\ref{FigShape}(a)) is not very suitable to represent the 'local'
spectral weight near $\omega_{i}$ \cite{DjurisicJAP01}. The
problem becomes evident from the following example. Imagine, that
there is a low-frequency gap $\omega_{g}$ in the spectrum:
$\epsilon_{2}(\omega)=0$ for $\omega<\omega_{g}$. However, each
Lorentzian that we have to put above $\omega_{g}$ is going to
contribute a non-zero value of $\epsilon_{2}$ below $\omega_{g}$
because of its low-frequency tail. This makes a good fitting
quality with Lorentzian oscillators hardly achievable.

Clearly, it is better to have a lineshape function, which is
non-zero only inside a small region, adjacent to $\omega_{i}$. A
good candidate is a triangular profile, shown on Figure
\ref{FigShape}(b):
\begin{equation}\label{TriagEps2}
    \epsilon_{2,i}^{\triangle}(\omega)=\left\{\begin{array}{cl}
      (\omega-\omega_{i-1})/(\omega_{i}-\omega_{i-1}) & \mbox{ , } \omega_{i-1}<\omega\leq\omega_{i} \\
      (\omega_{i+1}-\omega)/(\omega_{i+1}-\omega_{i}) & \mbox{ , } \omega_{i}<\omega<\omega_{i+1} \\
      0                                               & \mbox{ , } \mbox{otherwise} \\
    \end{array}\right..
\end{equation}

\noindent The real part can be obtained by the KK transformation
(\ref{kkeps1}):
\begin{eqnarray}\label{TriagEps1}
\epsilon_{1,i}^{\triangle}(\omega)=\frac{1}{\pi}\left[\frac{g(\omega,\omega_{i-1})}{\omega_{i}-\omega_{i-1}}
-\frac{(\omega_{i+1}-\omega_{i-1})g(\omega,\omega_{i})}{(\omega_{i}-\omega_{i-1})(\omega_{i+1}-\omega_{i})}\right.\nonumber\\
+\left.\frac{g(\omega,\omega_{i+1})}{\omega_{i+1}-\omega_{i}}\right]
\end{eqnarray}
\noindent with $g(x,y)\equiv(x+y)\ln|x+y|+(x-y)\ln|x-y|$.

Exactly as we did it before, the variational dielectric function
can be constructed as a linear superposition of triangular
functions sitting at fixed anchor frequencies $\omega_{i}$:
\begin{equation}\label{Super}
    \epsilon_{var}(\omega)=\sum_{i=2}^{N-1}A_{i}\epsilon_{i}^{\triangle}(\omega)
\end{equation}
\noindent The coefficients $A_{i}$ are tread as {\it free}
parameters. We excluded $i=1$ and $i=N$, from the summation by
defining $A_{1}=A_{N}=0$.

\begin{figure}[thb]
\includegraphics[width=7cm,clip=true]{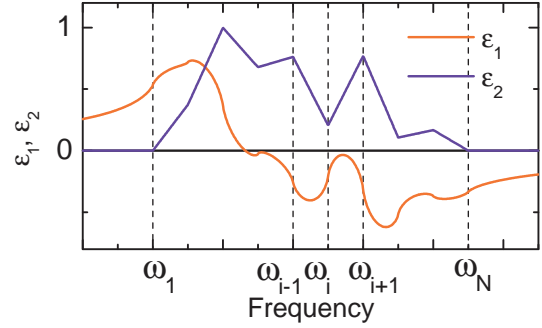}
\caption{An example of the variational dielectric function
$\epsilon_{var}(\omega)$, described in the text. The imaginary
part of $\epsilon_{var}$ is a superposition of several triangular
functions $\epsilon_{2,i}^{\triangle}(\omega)$, centered at anchor
frequencies. $\epsilon_{1,var}(\omega)$ is the exact KK transform
of $\epsilon_{2,var}(\omega)$.} \label{FigVar}
\end{figure}

The variational function is shown schematically on Figure
\ref{FigVar}. The imaginary part of $\epsilon_{var}(\omega)$ is a
piecewise curve passing through points $\{\omega_{i},A_{i}\}$ ($i
= 1\ldots N$). Thus, the adjustable parameters have very
straightforward meaning: they are the values of $\epsilon_{2}$ at
anchor frequencies $\omega_{i}$. Obviously, any shape of
$\epsilon_{2}(\omega)$ of a real material can be reproduced with a
proper set of parameters $A_{i}$, provided that the frequency mesh
is dense enough.

At first glance, a triangular lineshape is rather 'pathological'
for an oscillator. However, we are going to ascribe the physical
meaning not to the individual oscillators, but to the total
dielectric function (\ref{Super}), composed of many of them. It
makes us free to choose a lineshape, which is the most convenient
from the computational point of view. Even though this choice is
rather arbitrary, it is not crucial for the method. It is
equivalent to a decision of which quadrature formula to use for
the numerical integration in the KK transformation. The triangular
lineshape shape corresponds to the trapezoidal integration. One
can also use a rectangular shape of
$\epsilon_{2,i}^{\square}(\omega)$ (Figure \ref{FigShape}(c)).
However, it is inferior to $\epsilon_{2,i}^{\triangle}(\omega)$,
since the discontinuity of $\epsilon_{2,i}^{\square}(\omega)$
would result in artificial divergences of
$\epsilon_{1,i}^{\square}(\omega)$ at the rectangle edges.

\subsection{Discrete Kramers-Kronig transform}

Since we deal with the values of the dielectric function at a
fixed mesh, it is convenient to represent its real and imaginary
parts as vectors: $\epsilon_{1,i}\equiv \epsilon_{1}(\omega_{i})$
and $\epsilon_{2,i}\equiv \epsilon_{2}(\omega_{i})$. Then, using
formulas (\ref{Super}) and (\ref{TriagEps1}), the KK transform
(\ref{kkeps1}) can be represented in a matrix form:
\begin{equation}\label{KKdiscr1}
\epsilon_{1,i}-1= \sum_{j=1}^{N} K_{ij} \epsilon_{2,j},
\end{equation}
\noindent where
\begin{equation}\label{KKdiscr2}
K_{ij}=\epsilon_{1,j}^{\triangle}(\omega_{i}).
\end{equation}

\noindent Keeping in mind the analogy with the discrete Fourier
transform, we can call (\ref{KKdiscr1}) a {\it discrete
Kramers-Kronig transform} (DKKT). The matrix $K_{ij}$ is a
discrete analog of the KK kernel
$K(x,\omega)\equiv(2/\pi)x(x^2-\omega^2)^{-1}$ in (\ref{kkeps1}).
The exact expression for the DKKT depends, of course, on the
chosen oscillator lineshape. However, for $|i-j| \geq 2$ all
lineshapes give almost the same value of $K_{ij}$.

\subsection{The problem of extrapolations}

A serious shortcoming of the variational function (\ref{Super}),
taken alone, is that it does not take into account any spectral
weight beyond the frequency range $\omega_{1}<\omega<\omega_{N}$.
Due to the non-local character of the KK relation (\ref{kkeps1}),
the non-zero wings of $\epsilon_{2}$ at higher and lower
frequencies are essential to calculate $\epsilon_{1}$ inside this
region. Therefore, some assumptions about the high- and
low-frequency behavior of $\epsilon_{2}$ should be taken.

The same problem of extrapolations exists in the usual KK analysis
of reflectivity, which does not have any universal solution.
Nevertheless, an 'educated guess' can often be done, based on some
reasonable physical model. We argue that the following 'two-step'
fitting scheme can address the issue of the extrapolations.

At first, the experimental spectra can be fitted in a conventional
way by some formula-defined dielectric function
$\epsilon_{mod}(\omega)$ with a limited number of parameters. It
can be a DL function (\ref{DL}) or a more appropriate physical
model. If $\epsilon_{mod}(\omega)$ describes the major features of
experimental spectra then we can assume that it gives an
approximately correct frequency dependence outside the considered
spectral range, even though some fine spectral details may not be
fitted very well. Of course, in order to have a better confidence
in the extrapolations, the experimental spectral range has to be
expanded as much as possible, which is also true for the
conventional KK analysis of reflectivity.

The next step is to add the variational function (\ref{Super}):
\begin{equation}\label{comb}
\epsilon_{total}(\omega)=\epsilon_{mod}(\omega)+\epsilon_{var}(\omega),
\end{equation}
\noindent to 'fix' the parameters of $\epsilon_{mod}(\omega)$, and
to perform the final data fitting with only parameters $A_{i}$ of
the variational function (\ref{Super}) kept adjustable. The low-
and the high-frequency spectral weights are now accounted for by
$\epsilon_{mod}(\omega)$. In this case the 'flexible' function
$\epsilon_{var}(\omega)$ acts as a {\it small correction},
intended to fit the fine spectra details, unaccessible for the
'stiff' $\epsilon_{mod}(\omega)$. Therefore, the initial guess
values $A_{i}$ can be set to zero. Note that even though
$\epsilon_{2,total}(\omega)$ ought to be positive, the sign of
$\epsilon_{2,var}(\omega)$ is now arbitrary.

\subsection{Computational details}

Once the set of datapoints, the physical models and the anchor
mesh are specified, one can minimize the chi-square functional
(\ref{chisq}) using of the standard non-linear minimization
algorithms. In particular, the Levenberg-Marquardt (LM) method
\cite{NumericalRecipes} has proven to be very efficient, when the
number of parameters is large. An important advantage of the LM
algorithm is that it involves the calculation of derivatives of
the model spectra with respect to parameters using explicit
analytical formulas.

The optimization of the mesh of anchor frequencies
$\omega_{1}$,..,$\omega_{N}$ is rather important for the correct
work of the variational fitting routine. On one hand, the mesh
should be dense enough to enable the variational function
(\ref{Super}) to fit all important spectral details. On another
hand, $N$ should not be too large. The first reason is that the
calculation time grows quickly as a function of the number of
parameters to be adjusted. The second reason, discussed in Section
\ref{Problem}, is that an excessive 'flexibility' may cause
irregular behavior of the resulting functions, such as fast
spurious oscillations. According to our experience, for an optimal
mesh, $N$ is two times smaller, than $N_{exp}$, while every anchor
point corresponds to at least one experimental point, contributing
to formula (\ref{chisq}). On an everyday-use computer, one can
handle up to few thousands parameters within reasonable time
bounds, which is sufficient in most cases.

Another way to avoid physically meaningless numerical oscillations
would be to add extra 'regularization' terms to the total
chi-square functional (\ref{chisq}), intended to make unfavorable
those solutions, which have short-period oscillations. In this
paper we are not going to discuss the details of possible
regularization schemes. This work is in progress.

If the mesh optimization and other precautions fail to get rid of
a numerical instability, it may signal that the systematic error
bars are too large, or an inappropriate physical model is used.
Such 'inconsistency alert', which is missing in the usual KK
method, is another advantage of the proposed technique.

\section{Application examples}\label{Examples}

Now we turn to examples, showing the applications of the method.
All calculations were performed using Reffit software
\cite{RefFIT}, designed for the fitting of optical spectra with
various physical models as well as the KK-constrained variational
dielectric functions. As an experimental input, we shall use
either really measured curves, or computer-generated test spectra.

\subsection{Normal-incidence reflectivity}

\begin{figure}[thb]
\includegraphics[width=8.5cm,clip=true]{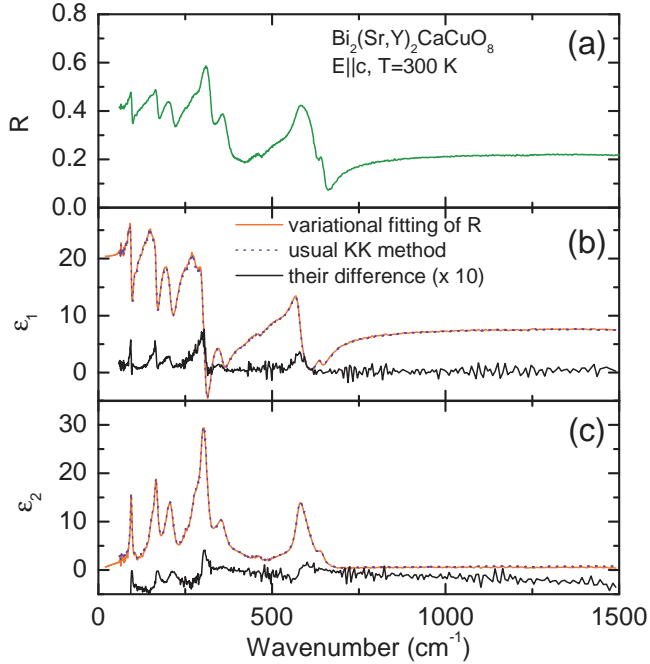}
\caption{The KK-constrained variational fitting vs. the usual KK
analysis, applied to the normal-incidence reflectivity of Bi-2212
for {\bf E} $\parallel$ {\bf c}, at room temperature (a). The two
methods give almost indistinguishable $\epsilon_{1}(\omega)$ (b)
and $\epsilon_{2}(\omega)$ (c). The difference between the
results, multiplied by 10, is shown for comparison.}
\label{FigBi2212}
\end{figure}

Let us first apply the variational fitting method to a
normal-incidence reflectivity spectrum and compare it with the
results of the usual KK analysis \cite{JahodaPRB57}. Figure
\ref{FigBi2212}(a) shows the far-infrared room-temperature
reflectivity of a high-$T_{c}$ superconductor
Bi$_{2}$(Sr,Y)$_{2}$CaCu$_{2}$O$_{8}$ (shortly Bi-2212) for the
electric field along the c-axis (perpendicular to the planes
CuO$_{2}$). The spectrum is dominated by optical phonons, since
the interplane electronic transport is almost blocked.

We applied the usual KK analysis, extrapolating the reflectivity
by a constant value at high and low frequencies, because the
conductivity of Bi2212 along the c-axis is very small.

In order to perform the variational fitting, we took a mesh of 700
anchor frequencies, logarithmically distributed between
$\omega_{1}$ = 60 cm$^{-1}$ and $\omega_{N}$ = 1500 cm$^{-1}$. The
resulting real and imaginary parts of $\epsilon(\omega)$ are shown
in Figure \ref{FigBi2212}(b) and (c). One can see that the two
techniques give almost identical results in this reference case,
when the conventional KK transform of reflectivity is known to be
operational.

\subsection{Grazing-incidence reflectivity}

The measuring of reflectivity $R_{p}(\omega)$ at a grazing angle
of incidence (close to 90$^{\circ}$) for the polarization of light
in the plane of incidence (p-polarization) is often preferable for
highly reflecting metallic samples, since the measured absorption
$A = 1 - R_{p}$ is more sensitive to the sample conductivity
\cite{SomalPRL96}, than it is for the normal reflectivity. We are
going to show that the variational fitting method is workable in
the case of the off-normal incidence reflectivity. Here we use an
artificial example, where the 'true' dielectric function is known
beforehand.

\begin{figure}[thb]
\includegraphics[width=8.5cm,clip=true]{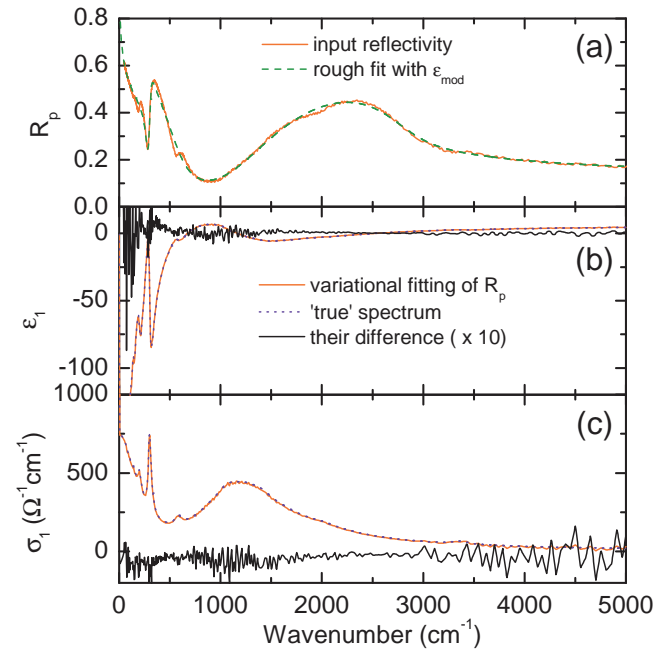}
\caption{The KK-constrained variational fitting of a
grazing-incidence reflectivity spectrum for $\theta$=80$^\circ$
(a, red solid curve), simulated, using 10 Lorentz and 1 Drude
oscillators as described in the text. The rough 3-oscillator fit
is shown by the dashed green curve. The resulting
$\epsilon_{1}(\omega)$ (b) and $\sigma_{1}(\omega)$ (c) are very
close to the 'true' spectra. The difference between the results,
multiplied by 10, is shown for comparison.} \label{FigGrazing}
\end{figure}

The 'true' dielectric function
$\epsilon(\omega)=\epsilon_{1}(\omega)+4\pi
i\sigma_{1}(\omega)/\omega$, which is shown  in Figure
\ref{FigGrazing} (b) and (c), was simulated by a DL model
(\ref{DL}) with 10 randomly put Lorentz oscillators, one Drude
peak and $\epsilon_{\infty}$=6. The 'experimental' reflectivity
for the p-polarized light for the angle of incidence
$\theta$=80$^{\circ}$ was calculated by the Fresnel
formula\cite{DresselGruner}:
\begin{equation}
R_{p}=\left|\frac{\epsilon\cos\theta-\sqrt{\epsilon-\sin^{2}\theta}}
{\epsilon\cos\theta+\sqrt{\epsilon-\sin^{2}\theta}}\right|^{2}.\nonumber
\end{equation}
Some statistical noise of 1\% was added (red solid line on Figure
\ref{FigGrazing} (a)) in order to mimic a real-life situation. We
verified that the usual KK analysis does not work in this case:
the formula (\ref{kkr}), when applied to $R_{p}$, gives a totally
wrong reflectivity phase spectrum.

The obtained in this way $R_{p}(\omega)$ was taken as an input for
the KK-constrained variational fitting procedure. At the first
stage the rough fitting was performed with only 3 oscillators
included to $\epsilon_{mod}(\omega)$ (green dashed line on Figure
\ref{FigGrazing} (a)). After that a variational function
$\epsilon_{var}(\omega)$ was added to $\epsilon_{mod}(\omega)$ and
the fine fitting of $R_{p}(\omega)$ was done. We took a mesh of
500 anchor frequency points, logarithmically distributed between
50 cm$^{-1}$ and 10000 cm$^{-1}$. The resulting
$\epsilon_{1}(\omega)$ and $\sigma_{1}(\omega)$ (Figures
\ref{FigGrazing} (b) and (c), red line) are very close (apart from
the noise) to the 'true' ones (Figure \ref{FigGrazing} (b) and
(c), blue line). Of course, such a good agreement is only
possible, when the frequency range of the input reflectivity
spectrum is large enough.

An example of the application of this procedure to a really
measured grazing-incidence reflectivity spectrum (in combination
with ellipsometry at higher frequencies) can be found in
Ref.\cite{MenaPRB03}.

\subsection{Reflectivity + transmission}

\begin{figure}[thb]
\includegraphics[width=8.5cm,clip=true]{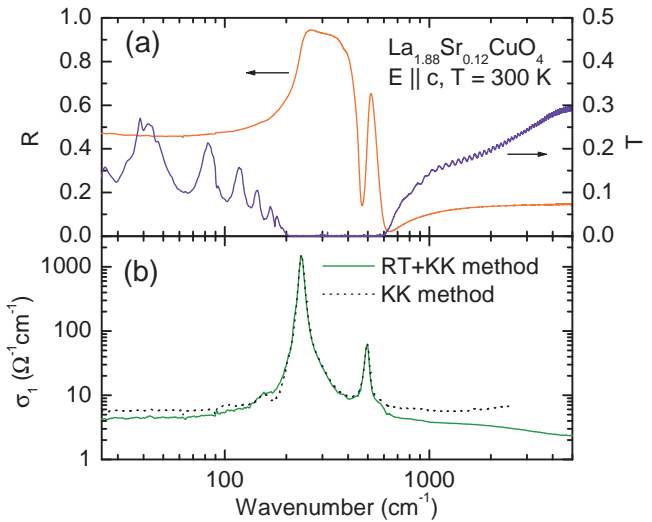}
\caption{Reflectivity (bulk sample) and transmission (thin 23
$\mu$m sample) of LSCO for E$\parallel$c at room temperature. (b)
The conductivity spectrum obtained by the RT+KK analysis, as
described in the text.} \label{FigLSCO}
\end{figure}

Now we consider an example, when two measured spectra {\it of
different types} are analyzed simultaneously. Figure
\ref{FigLSCO}(a) shows reflectivity and transmission spectra of
La$_{1.88}$Sr$_{0.12}$CuO$_{4}$ (LSCO) in the far- and
mid-infrared range at room temperature \cite{KuzmenkoPRL03}. The
electric field was always parallel to the c-axis. The near-normal
incidence reflectivity $R(\omega)$ was measured on a thick
non-transparent sample. The transmission spectrum $T(\omega)$ was
taken on a thin slab of the same material with a thickness $d$=23
$\mu$m. Both spectra can be expressed in terms of the complex
dielectric function $\epsilon(\omega)$ \cite{DresselGruner}:
\begin{equation}\label{t}
R=|r|^2, T=\left|\frac{(1-r^{2})t}{1-r^{2}t^{2}}\right|^2,
\end{equation}
\noindent where
\begin{equation}
r=\frac{1-\sqrt{\epsilon}}{1+\sqrt{\epsilon}}, t=\exp
\left(i\frac{\omega}{c}\sqrt{\epsilon}d\right).\nonumber
\end{equation}
\noindent The formula for $T$ takes into account multiple internal
reflections, which give rise to Fabry-Perot interference fringes.
Unfortunately, the fringes are partially suppressed in the
experiment, because the sample was not exactly flat-parallel. This
was taken into account by the averaging of $T$ for a certain
distribution (about 10\%) of the sample thicknesses.

Figure \ref{FigLSCO}(b) shows the optical conductivity
$\sigma_{1}(\omega)$, obtained by two different techniques: the
standard KK transformation of reflectivity (not considering the
transmission spectrum) and the KK constrained variational
simultaneous fitting of reflectivity and transmission (which we
call RT+KK method\cite{KuzmenkoPRL03}). The c-axis conductivity is
featured by a slowly varying small electronic conductivity,
determined by an incoherent electron transport along the c-axis,
and the two strong phonon peaks at 235 and 495 cm$^{-1}$.

One can notice, that the two techniques give similar results in
the phonon range, but significantly deviate at lower and
especially at higher frequencies (watch the log scale!), where the
conductivity is small. It is not surprising, since the
reflectivity in this case depends mostly on the $\epsilon_{1}$ and
is not very sensitive to $\sigma_{1}$. Thus, the error bars of the
latter are very large. On the other hand, the transmission is
directly related to the absorption, and, correspondingly, to the
conductivity. Therefore, it is really worthwhile to use both
$R(\omega)$ and $T(\omega)$ in the analysis.

The advantage of using the KK constraint here is not immediately
obvious, since the two values $R$ and $T$ can be in principle
converted to $\epsilon_{1}$ and $\sigma_{1}$ at every given
frequency (RT-method). However, this conversion works well only if
the value of transmission is not too small. In our case the
transmission vanishes between 200 and 600 cm$^{-1}$ (in the phonon
range), where we have to resort to the KK relations. Using two
different techniques in different ranges would not be very
practical. Alternatively, the RT+KK method provides a seamless
conductivity spectrum with small error bars in the whole spectral
range. In addition, the using of the transmission spectrum outside
the phonon range in the RT+KK technique suppresses the
uncertainties of the KK analysis due to the high- and
low-frequency extrapolations. A similar idea of 'phase anchoring'
is discussed in the next example.

\subsection{Low-frequency reflectivity + high-frequency $\epsilon_{1}$ and $\epsilon_{2}$}

It is known, that the accuracy of the KK analysis can be greatly
improved if both real and imaginary parts of $\epsilon(\omega)$
are measured {\it independently} in selected frequency points
\cite{VartiainenAS96,MiltonPRL97} or in a certain interval
\cite{HulthenJOSA82,BozovicPRB90}. Here we consider a situation,
when one can obtain $\epsilon_{1}(\omega)$ and
$\epsilon_{2}(\omega)$ at high frequencies, but only the
reflectivity $R(\omega)$ can be measured at low frequencies.

In order to extract the conductivity at low frequencies, we have
to use the KK relations. It is desirable to take as much profit as
possible from the knowledge of both $\epsilon_{1}$ and
$\epsilon_{2}$ at high frequencies. Therefore, it would be rather
wasteful, for example, to calculate $R$ from $\epsilon_{1}$ and
$\epsilon_{2}$ at high frequencies and use it only to extend the
range for the usual KK transformation (\ref{kkr}). On the other
hand, the simultaneous KK-constrained fitting of the low-frequency
reflectivity and the high-frequency $\epsilon_{1}(\omega)$ and
$\epsilon_{2}(\omega)$ \cite{EbPRL01,KuzmenkoSSC02,MenaPRB03}
allows one to use all available information, in particular, to
'anchor' the phase of the reflectivity.

\begin{figure}[thb]
\includegraphics[width=8.5cm,clip=true]{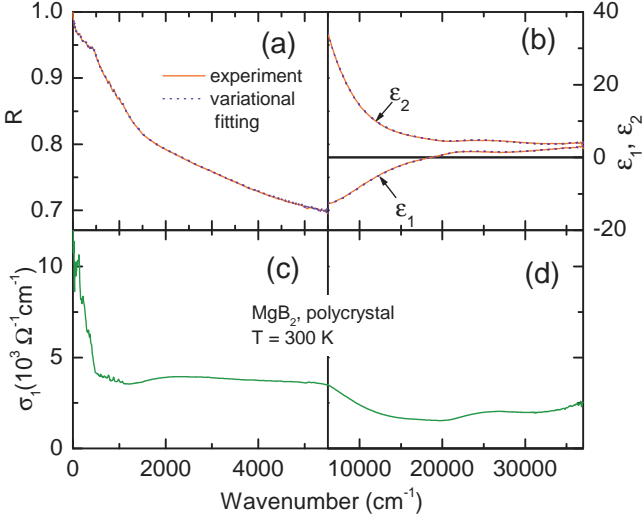}
\caption{The KK-constrained variational fitting of the
reflectivity in the far- and mid-IR (a) and $\epsilon_{1}(\omega)$
and $\epsilon_{2}(\omega)$ in the near-infrared to UV ranges of a
polycrystalline sample of MgB$_{2}$ at room temperature. The
panels (c) and (d) show the obtained effective optical
conductivity.} \label{FigMgB2}
\end{figure}

An example is shown on Figure \ref{FigMgB2}. The data were taken
on a polycrystalline sample of MgB$_{2}$, which is a good metal.
Even though the magnesium diboride is an anisotropic compound, in
the spectra analysis it was assumed that the powder sample behaves
as an optically isotropic one, described by an effective
dielectric function $\epsilon(\omega)$. The reflectivity at a
nearly normal incidence was measured in the range 20 - 6000
cm$^{-1}$ (Figure \ref{FigMgB2}(a)), and the ellipsometry at an
angle of incidence $\theta$=80$^{\circ}$ provided
$\epsilon_{1}(\omega)$ and $\epsilon_{2}(\omega)$ at 6000-37000
cm$^{-1}$ (Figure \ref{FigMgB2}(b)).

The variational fitting was done in the same way, as described
before. The anchor mesh covered both spectral ranges. The quality
of the variational fit is very good, giving confidence in the
resulting optical conductivity spectrum (Figure
\ref{FigMgB2}(c-d)).

\subsection{Reflectivity of a thin film on a substrate}

The next example shows that the method can also be used to analyze
optical data of layered samples. We shall consider a simple but
practically very important case of an optically thin film grown on
a substrate. If the penetration depth is comparable, or larger
than the film thickness then the substrate contributes to the
overall optical response. One has to take it into account, when
extracting optical properties of the film from the reflectivity
spectra \cite{SantanderPRL02}.

\begin{figure}[thb]
\includegraphics[width=8.5cm,clip=true]{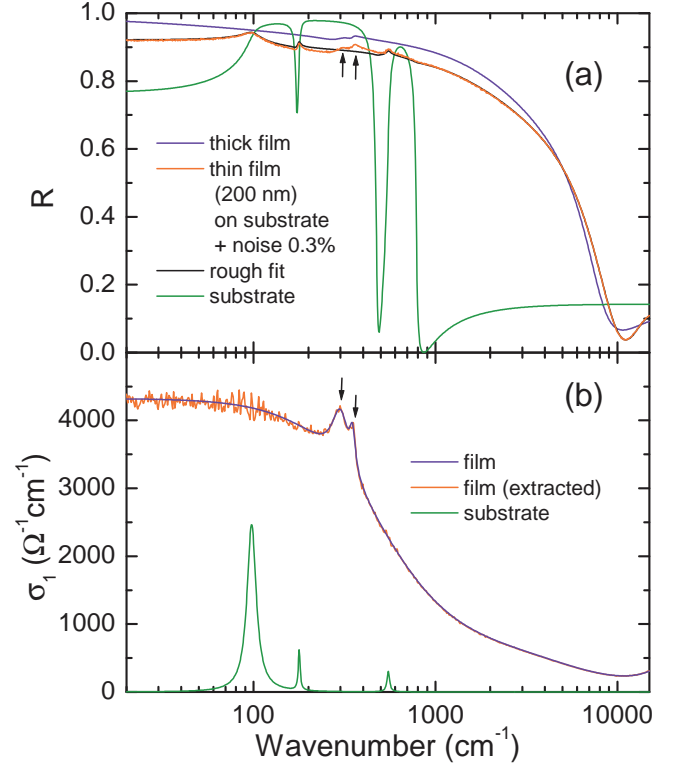}
\caption{A test example of the KK-constrained variational fitting
of a reflectivity spectrum measured on a thin film (200 nm) grown
on a substrate with known optical properties. The test dielectric
functions of the film and the substrate were generated as
described in the text. The panel (a) shows the generated
reflectivity of a thick film (blue), the one of the thin film
(red), used as an input, with a noise of 0.3\% added, and the one
of a bare substrate (green). The black curve  is  the initial
rough fit of the 'measured' reflectivity. The panel (b) indicates
the exact optical conductivity of the film (blue), the extracted
one by the method (red) and the conductivity of the substrate
itself (green). On both panels the arrows mark the optical peaks,
which come from the film. } \label{FigThinFilm}
\end{figure}

We generate dielectric functions of the film
$\epsilon_{f}(\omega)$ and the substrate $\epsilon_{s}(\omega)$ in
the infrared range using certain models in order to check later on
the accuracy of the final result. For the substrate we took a
model that gives optical properties close to the ones of
SrTiO$_{3}$ \cite{SpitzerPR62}, often used as a substrate
material. The reflectivity $R(\omega)$ and optical conductivity
$\sigma_{1}(\omega)$ of the substrate are shown by green curves in
Fig. \ref{FigThinFilm} (a) and (b) respectively. The infrared
spectrum of SrTiO$_{3}$ is dominated by strong phonon lines at
100, 180 and 550 cm$^{-1}$. To simulate the properties of the
film, a model was taken, which gives a metallic conductivity with
a plasma edge at around 10000 cm$^{-1}$ and some additional
double-peak structure at about 300 - 400 cm$^{-1}$. The
corresponding $R(\omega)$ (for a bulk sample) and
$\sigma_{1}(\omega)$ are presented in Fig. \ref{FigThinFilm} (blue
curves).

The normal-incidence 'experimental' reflectivity of the
film+substrate system was calculated using formula
\cite{DresselGruner}
\begin{equation}
R=\left|\frac{r_{f}+t^2r_{fs}}{1+t^2r_{f}r_{fs}}\right|^2 ,
\end{equation}
\noindent where
\begin{equation}
r_{f}=\frac{1-\sqrt{\epsilon_{f}}}{1+\sqrt{\epsilon_{f}}},
r_{fs}=\frac{\sqrt{\epsilon_{f}}-\sqrt{\epsilon_{s}}}{\sqrt{\epsilon_{f}}+\sqrt{\epsilon_{s}}},
t=\exp\left(i\frac{\omega}{c}\sqrt{\epsilon_{f}}d\right)\nonumber
\end{equation}
\noindent for the film thickness $d$ of 200 nm (the red curve in
Fig.\ref{FigThinFilm}(a)). Some artificial noise of 0.3\% was
added, in order to check the influence of the statistical errors
on the result. One can see that the reflectivity spectrum of the
thin films is very different form the one of a thick sample. The
structures in the far-infrared range are caused not only by peaks
in the film conductivity (shown by arrows), but also by the phonon
lines of the substrate.

As usual, the fitting process was done in two steps. Initially,
the 'experimental' spectrum was fitted with a very limited model
$\epsilon_{mod}(\omega)$. The corresponding reflectivity is shown
by the black curve in Fig.\ref{FigThinFilm}(a). Secondly, the
variational fitting was performed, using $\epsilon_{var}$, defined
on about 500 anchor frequency nodes (one-half of the total number
of experimental points).

The resulting optical conductivity is shown by the red curve in
Fig.\ref{FigThinFilm}(b). The agreement with the exact
conductivity is rather good. The double-peak structure, which was
not described by the initial limited model, is reproduced quite
well. One can see that the experimental noise has little effect at
high frequencies, but is strongly amplified below 200 cm$^{-1}$.
This is hardly surprising, since the KK method is known to produce
large error bars for metallic samples at low frequencies, where
the reflectivity is close to 1.

\section{Summary}\label{Summary}

We introduced a new technique of the extraction of the real and
the imaginary parts of the dielectric function from optical
spectra. The idea is that all available datapoints are fitted
simultaneously, using a dielectric function, which has a fully
flexible (variational) imaginary part $\epsilon_{2}(\omega)$ and
the real part $\epsilon_{1}(\omega)$, dynamically coupled to the
$\epsilon_{2}(\omega)$ via the KK relation. In simple cases the
method works as well, as the usual KK transformation of
reflectivity.

It was shown that the scope of applications of the variational
KK-constrained fitting is extremely broad. It extends to various
optical techniques, or combinations of them, as well as to
different types of samples. The computational time required for a
typical KK-constrained fitting session is somewhat longer,
compared to the usual KK transformation, but it is yet well within
reasonable limits. It is important, that the method needs almost
no readaptation for different experimental situations, saving a
lot of human time.

Certain optical methods, such as spectroscopic ellipsometry, or a
combination of reflection and transmission do not formally rely
the KK relations, since both $\epsilon_{1}$ and $\epsilon_{2}$ can
be obtained directly from the measured data at each frequency.
Even though the independence on the KK relation is a big advantage
of such techniques, the method, proposed in this paper, can yet be
very useful here as a stringent test on the overall KK consistency
of the experimental data.

In this article we assumed that the materials under study are
non-magnetic, which means that magnetic permeability $\mu$ is
equal to 1. For magnetic materials, the method can be used to
determine $\mu(\omega)$, which also satisfies the KK relations. Of
course, in order to disentangle $\mu(\omega)$ and
$\epsilon_(\omega)$ one may need to analyze several measurements
simultaneously (for instance, the transmission of samples of
different thicknesses).

So far we discussed the linear-response optical measurements only,
but, of course, the range of subjects, where the KK relations are
used, is not restricted to the linear optics. The method is
potentially applicable to any spectroscopic technique, where the
measured spectra depend on a combination of real and imaginary
parts of a causal, a therefore KK-compliant, response function to
be determined. The examples include photoemission, neutron
scattering{\cite{KuzmanyBook}}, X-ray spectroscopy
\cite{KramersNature26,KronigJOSA26,BonseNIMPR84}, electron energy
loss spectroscopy (EELS)\cite{WhiteJESRP87}, magneto optical Kerr
effect (MOKE) \cite{KielarJOSB94}, Josephson-contact
spectroscopy\cite{ZwergerSSC83,FerrellPC88}, electrode impedance
measurements\cite{MeirhaegheEA75}, electron spin resonance
(ESR)\cite{AltshulerSPSS73, JaroszPSSB81}, low-energy electron
diffraction (LEED)\cite{KinniburghJPC78}, acoustics
\cite{ZelloufJAP96,MobleyJASA03} and others. Since KK-type
relations exist for higher-order susceptibilities
\cite{YakubovichJETP69,BassaniPRB91}, this way to analyze data can
also be useful in non-linear spectroscopic techniques.

The author is grateful to D. van der Marel, F. P. Mena, H. J. A.
Molegraaf and F. Carbone for numerous discussions and invaluable
practical help. Useful advices about solutions of ill-posed
inverse problems were given by S. V. Rotin.


\begin{references}
\bibitem{Text1}
Here we assume that the magnetic permeability $\mu=1$ and the
effects of the spatial dispersion are negligible, or, in other
words, $\epsilon(\omega)$ can be regarded as momentum-independent.
\end{references}
\end{document}